\documentstyle[10pt,aas2pp4]{article}

\lefthead{Explosion diagnostics of Type Ia supernovae from early infrared}
\righthead{Explosion diagnostics of Type Ia supernovae from early infrared}

\begin{document}
\title {Explosion Diagnostics of Type Ia Supernovae
from Early Infrared Spectra}
\author{J. Craig Wheeler, Peter H\"{o}flich, Robert P. Harkness}
\affil{Department of Astronomy, University of Texas, Austin, TX78712, USA}
\and
\author{Jason Spyromilio}
\affil{European Southern Observatory, Karl-Schwarzschild-Strasse 2, D-85748,
Garching bei M\"unchen, Germany }

\begin{abstract}

Models of infrared spectra of Type Ia supernovae around maximum light
are presented. The underlying dynamic models are delayed detonation
explosions in Chandrasekhar mass carbon/oxygen white dwarfs.  In
combination with the radiative transport codes employed here, these
models provide plausible fits to the optical spectra of ``normal"
Type Ia supernova.  Two independent radiative transport codes are used,
one that assumes LTE and one that computes non-LTE excitations and
ionization. The models are compared with infrared data available in the
literature. The independent codes give a reasonable representation of
the data and provide physical explanations for their origin independent
of the detailed assumptions of the radiative transfer.  The infrared
gives an especially powerful diagnostic of the dynamic model because
with strongly variable line blanketing opacity it
probes different depths within the exploded white dwarf at the same
epoch. The velocity of the transition zone between explosive oxygen and carbon
burning  can be directly determined. The
velocity at which the burning to nickel stops can also be probed. These
velocities are very sensitive to the explosion physics.

\end{abstract}


\section{Introduction}

Infrared spectroscopy of supernovae at early times has been very
sparse, mostly due to the limited number of facilities equipped with
the appropriate instrumentation.  As a result, the usefulness of such
spectra as diagnostics of supernovae has not been investigated fully.
The lack of models to appropriately explore these data has, in some
cases, discouraged their publication.  The lack of published data, in
turn, has  resulted in model spectra being artificially truncated at a
wavelength of 1 $\mu$m.  Here we will concentrate on interpretation
of the IR spectral data of Type Ia supernovae (SN~Ia).

Kirshner et al. (1973) and Branch et al. (1983) showed that after maximum
in SN~Ia the flux in J is depressed with respect to the continuum at longer
and shorter wavelengths.
Subsequent IR spectroscopy revealed a pronounced broad spectral
minimum in the J band at $\sim$ 1.2 $\mu$m (Frogel et al.
1987, Meikle et al. 1996) and other spectral features.
Spectroscopic analysis of this early time data is limited.
Meikle et al.  (1996) concentrated on
determining the origin of a feature at 1.05 $\mu$m in the spectra of
SN\,1994D. They concluded that the feature might be due to either HeI 1.083
$\mu$m
or MgII 1.0926  $\mu$m, but found difficulties with both identifications.
Lynch et al. (1990) were unable to determine a
plausible explanation for the 1.2 $\mu$m feature observed in a
spectrum of SN\,1989B.
Graham (1986) attributed the 1.2 $\mu$m spectroscopic feature to absorption by
a multiplet of SiI.  H\"oflich et al. (1993) showed that the line
opacity has a deficit in the 1.2 $\mu$m range but did not discuss the
possible implications for the 1.2 $\mu$m spectral feature.
Spyromilio, Pinto \& Eastman (1994) argued that no
multiplet or combination of lines is responsible for the lack of flux
in the 1.2 $\mu$m spectral region.   Instead, it is the absence of a line
blanketing opacity and the relative transparency of the supernova in
continuum opacity that causes the
lack of flux in the 1.2 $\mu$m region.  The work we report here supports
the interpretation that a lack of line blanketing contributes to the paucity of
flux at J with respect to H and puts that interpretation in a broader
context.

With the interpretation that some portions of the IR have relatively large
line blanketing and others relatively little,
the infrared region of the spectrum provides
a unique window into the supernova. In adjacent wavelength regions the
supernova ejecta can be probed both deeply within the ejecta where a
paucity of lines provides a minimum in the opacity and at shallow depths where
strong line blanketing provides a quasi-continuum formed at
relatively large radii.
In this paper, we explore the formation and evolution of the spectra
of SN Ia throughout the wavelength range from 0.9 to 2.5 $\mu$m from about a
week before to about two weeks after maximum light. We also elucidate the
relevant physical diagnostics at different wavelengths and phases.
In \S2 we describe some observations
available to us which provide a basis for comparison of the models.  In
\S3 we describe the models used to probe the physics.  In \S4
we confront the theoretical results with observations.  It is
important to note that we are not attempting to provide a best fit to
the  limited available data.  The objects for which we have some data are not
necessarily garden variety SN Ia.  In \S5, we
provide a final discussion and conclusions.

\section{Observations}

A limited number of existing early time IR spectra exist.  The data
reproduced here have been taken from the literature (Meikle et al. 1996;
Bowers et al. 1997).

The data for SN\,1986G were provided by
Graham (private communication).  SN\,1986G exploded in the dust lane of
Centaurus A and exhibited bluer infrared colors and shorter timescales
for spectral variations than a ``normal" SN Ia (Frogel et al.
1987).  The spectrum shown in Figure\,1 was obtained at the UKIRT 3.8-m
telescope on three consecutive  nights in 1986, May 23, 24 and 25.  The
B maximum light occurred on 1986, May 11$\pm$1 and the V maximum light was
attained 2-3 days later (Phillips et al.  1987).

\begin{figure}[t]
\centering
\plotfiddle{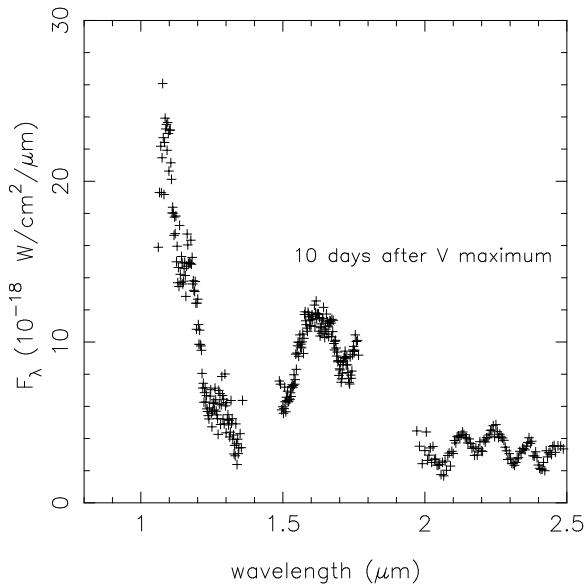}{7cm}{0}{45}{45}{-120}{0}
\caption{Spectrum of SN 1986G 10\,d after visual maximum}
\label{ Ia supernovae at early times}
\end{figure}

SN\,1994D is the most normal SN Ia for which good IR data
exist (Meikle et al. 1996).  Although pre-maximum light data are
available in all the near-infrared bands, only the $J$-band was
followed significantly past the first epoch of observation.  The first
spectrum was obtained about 8.5 days before B maximum.

\section{Description of the Numerical Methods}

\subsection { Hydrodynamics}

The explosions are calculated using a one-dimensional radiation-hydro
code, including nuclear networks (H\"oflich \& Khokhlov 1996, and
references therein).  This code solves the hydrodynamical equations
explicitly by the piecewise parabolic method (Collela \& Woodward 1984)
and includes the solution of the frequency-averaged radiation transport
implicitly via moment equations, expansion opacities, and a detailed
equation of state. The frequency-averaged variable Eddington factors
and  mean opacities are calculated by solving the frequency-dependent
transport equations.
About one thousand frequencies (in one hundred frequency groups) and
about five hundred depth points are used.  Nuclear burning is
taken into account using a network which has been tested in many
explosive environments (see Thielemann, Nomoto \& Hashimoto 1996, and
references therein).

\subsection {Spectral Calculations}

Both the LTE (Wheeler \& Harkness 1990) and non-LTE (H\"oflich, Wheeler
\& Thielemann 1997 and references therein) codes solve the relativistic
radiation transport equations in a comoving frame.
Neither code fixes the luminosity of the supernova using the
so-called ``light bulb" approach.  Rather, the energetics of the model
are calculated. Given an explosion model, the evolution of the spectrum
is not subject to any
tuning or free parameters.

The non-LTE spectra are computed for various epochs using the chemical,
density and luminosity structure and $\gamma$-ray deposition resulting
from the light curve code (H\"oflich, Wheeler, \& Thielemann 1997, and
references therein), thus providing a tight coupling between the explosion
model and the radiative transfer.  The effects of instantaneous energy
deposition by $\gamma$-rays, the stored energy (in the thermal bath and
in ionization) and the energy loss due to the adiabatic expansion are
taken into account.  The radiation transport equations are solved
consistently with the statistical equations and ionization due to
$\gamma$-radiation for the most important elements (C, O, Ne, Na, Mg,
Si, S, Ca, Fe, Co, Ni).  About 10$^{6}$ additional lines are included
assuming LTE-level populations.
 The scattering, photon redistribution, and thermalization terms
are computed with an equivalent-two-level formalism that is calibrated using
NLTE
models (H\"oflich, 1995).
Typically, the photon redistribution and thermalization terms are
$\sim$ $10^{-2}$ to $10^{-3}$ in the optical and $\sim$0.1 in the IR.

 The LTE spectra are computed using similar input
models and also using a consistent energy deposition calculation. The
effects of instantaneous energy deposition, stored energy, and the
energy loss due to the expansion are all taken into account
using the approximation of flux limited diffusion.  The main
differences between the two codes are the energy level populations and
the treatment of line scattering, photon redistribution and
thermalization. For the LTE calculations, the thermalization parameter
is assumed to be 0.01, independent of frequency.

\subsection{General Results}

Explosion models have been taken from the literature.  No specific
tuning of the input explosion models has been made to produce a best
fit to the data and the radiation transport codes were run blind.  No
knowledge of the data was available at the time of the code execution.
Slightly different explosion models have been used in the two spectral
synthesis calculations.  For the LTE code model DD21 (H\"{o}flich,
Wheeler \& Thielemann 1997) was used while for the non-LTE calculation
model DD200 (H\"{o}flich, Khokhlov \& Wheeler 1997) was used.  Both are
delayed detonation models.  Investigation of other explosion models
lies outside the scope of this work.  In DD21 the central density of
the C/O white dwarf and the density of transition from deflagration to
detonation were $\approx $ 20 percent higher than in DD200, making DD21 a bit
more
luminous due to a higher $^{56}$Ni production.  Figures 2 and 3 show
the composition structure in velocity space for DD200 and DD21. The
velocities of the intermediate mass elements, especially silicon, are
somewhat different in the two models.

\begin{figure}
 \centering
 \plotfiddle{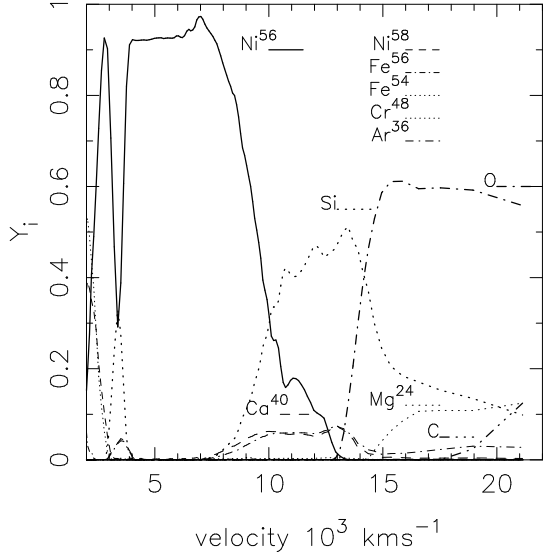}{8cm}{0}{45}{45}{-120}{0}
 \caption{The composition of DD200}
 \label{composition of DD200}
\end{figure}

\begin{figure}
 \centering
 \plotfiddle{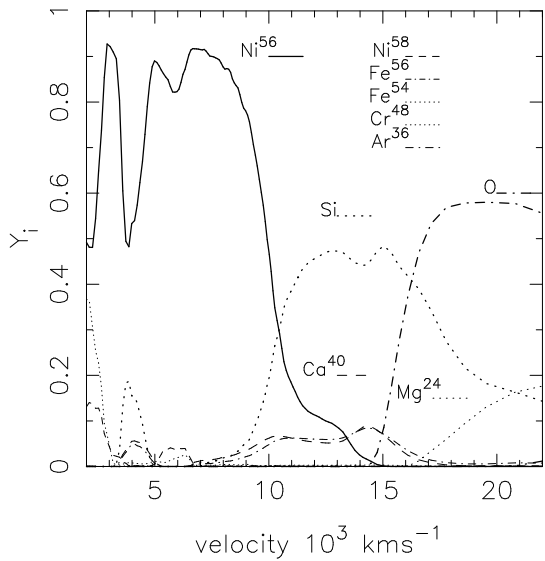}{8cm}{0}{45}{45}{-120}{0}
 \caption{The composition of DD21}
 \label{composition of DD21}
\end{figure}

\begin{figure}
 \centering
 \plotfiddle{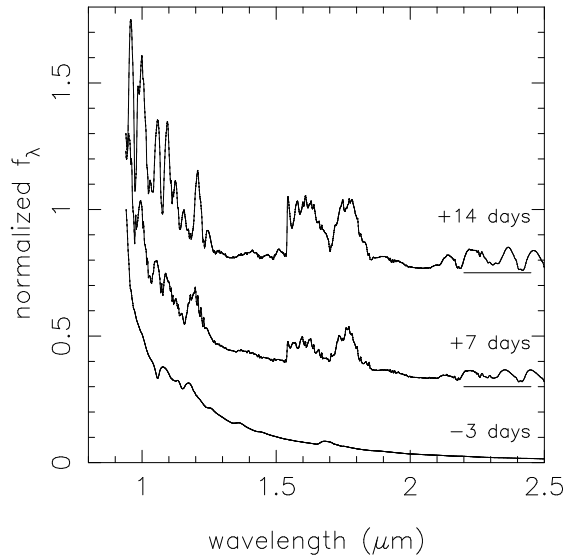}{8cm}{0}{45}{45}{-120}{0}
 \caption{Evolution of the DD200 model spectrum (times with respect
to model V maximum)}
\end{figure}

Figure\,4 gives the evolution of the spectrum for DD200.  The epochs
are given with respect to model V maximum.  They are -3, +7 and +14 days
corresponding to about 15, 25, and 32 days after the explosion.  The spectra
have been normalized to the maximum flux in the displayed range and then a
constant has been added for clarity of display.  The levels of zero
flux for the spectra after maximum light are shown as straight lines in
the red part of the spectrum.

Figure\,5 presents the evolution of the radius at which of the optical
depth is unity for the non-LTE model.  The wavelength attributed to
each radius is that of the principle ray along the line of sight to the
observer and therefore the wavelengths of the radius peaks are by
default blue shifted with respect to the corresponding peaks in
the spectrum.  The plots for +7 and +14 d have been shifted upward
for clarity.  The levels of zero radius for those epochs are shown
by the straight lines on the right side of the diagram.

Three days before V maximum, the spectrum in Figure\,4 is relatively
featureless with the exception of Mg\,II, Ca\,II and
Si\,II lines at 1.05, 1.12 and 1.65 $\mu$m, respectively.  The
opacities are dominated by electron scattering.  Later on, at
wavelengths shorter than $\sim$1.2 $\mu$m, the spectra are dominated by
blends of Mg, Ca and iron group elements.  The change of the slope of
this region with time provides a good test for the epoch of the
observed supernova spectrum.  With increasing time, the atmosphere
becomes cooler. Once the electron scattering opacity decreases
substantially, a SN Ia no longer has a well defined
photosphere (Wheeler \& Harkness 1990, Jeffery et al. 1992, H\"oflich
et al. 1993, Spyromilio, Pinto \& Eastman 1994).  The depth of the photosphere
varies with wavelength depending on the opacity of transitions present in that
frequency interval (Fig.\,5).  The features at about 1.5---1.7 and
2.2---2.6 $\mu$m are primarily due to iron group elements of the second
ionization stage with a dominant contribution from Co and a substantial
contribution from Ni.  The features at 2.2---2.6 $\mu$m also contain
appreciable contributions from intermediate mass elements, especially
Si.  The small bump in the spectrum corresponding to  3 days before maximum at
1.7 $\mu$m in Figure\,4 is from Si\,II.  These features appear in
``emission" because they are formed at a larger radius with corresponding
larger effective area.  In addition, the source
function is not scattering dominated, but governed by the
redistribution of energy from higher frequencies (H\"oflich 1995).
In contrast to the optical, in the IR photons are trapped
within the broad opacity bands that keep the mean photon density
and, thus the source function, higher than simple geometrical dilution of the
radiation field would give.

Flux minima occur where the spectrum is formed at a small radius
(Fig.\,5).  Again, a principle effect is
that the emitting area is small.  At -3 days the total continuum optical
depth is about 30.  By 14 days after maximum light the continuum has decreased
substantially.  At this time, electron scattering, free-free and bound-free
opacity give an optical depth in the continuum of $\sim$ 3.  At 1.2 $\mu$m
line blanketing is only about 20 percent of the total opacity, whereas
line blanketing dominates the opacity at 1.6 and 1.8 $\mu$.

  The peaks and valleys in the opacity distribution may be used as a
tool for analyzing the composition of the ejecta.
The broad holes between 1.2 and 1.5 and 1.9 and 2.1 $\mu$m correspond to a
dispersion in velocity of 70,000 and 30,000\,km\,s$^{-1}$, far
exceeding the dispersion of the expansion velocity of radioactive
material in any supernovae.  Consequently, these gaps cannot be
eliminated by velocity smearing and should in general be visible when
the continuum opacity is low enough to expose them.  The existence of
the gaps provides the opportunity to probe very different layers of the
ejecta at a given time.  This makes the IR an important complement to
the optical and the UV.

As mentioned above, the peaks at about 1.6 and 1.8 $\mu$m are due to
iron peak elements, specifically lines of Ni and Co.  They thus sample
the conditions of the radioactive elements.  The narrow minimum seen at
1.7 $\mu$m is due to a lack of strong blended lines in this wavelength
range. Both the blue and red edges of this feature are formed in layers
expanding
towards the observer and, consequently, blueshifted according to the maximum
velocity of the radioactive material. This flux minimum can be smeared
out by velocity shifts of about 10,000\,km\,s$^{-1}$. Such shifts are
comparable with the expansion velocity of the radioactive material,
therefore this narrow gap is a valuable diagnostic tool for the
velocity spread of the region containing the radioactive material in
which the feature is formed.  Because the envelope becomes more
transparent with increasing time exposing a broader range of
velocities, the velocity spread of the visible
iron group elements increases.  This causes the 1.7 $\mu$m gap to
becomes narrower with time (Fig.\,4).  Clearly, the time evolution and
the strength of the 1.7 $\mu$m gap will depend sensitively on the
underlying model.

\begin{figure}
 \centering
 \plotfiddle{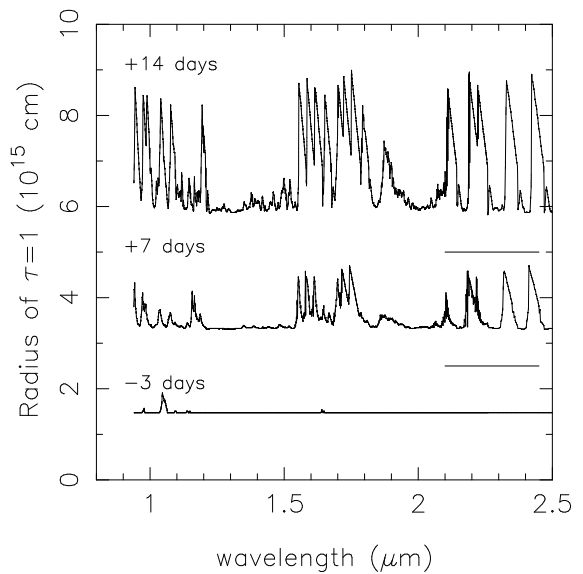}{8cm}{0}{45}{45}{-120}{0}
 \caption{Evolution of the DD200 radius of optical depth unity}
\end{figure}

\section{Comparison with the Observations}

We neither tuned the models to provide a best fit nor reproduced the
exact times of the observations.  Instead, the goal of this section is
to demonstrate that the IR is a valuable tool to analyze SN Ia
and to test whether our models can provide a reasonable
explanation for the observations.

\subsection{Pre-maximum Spectra}

The premaximum data on SN\,1994D are compared with the pre-maximum
non-LTE models for DD200 in Figure\,6.  The data were obtained 8.5\,d
before B maximum light while the theoretical spectrum is 2 and 3\,d
before B and V maximum light, respectively.  In this model, the strong
feature in the model at 1.05 $\mu$m is due to Mg\,II
and not to the helium triplet transition.
The He abundances in DD200 and DD21c are rather typical for delayed detonations
and the deflagration model W7.  Neither
has sufficient helium in the ejecta to form the line.
Moreover, magnesium is a natural product
of the explosive nucleosynthesis and therefore a more plausible
candidate.  As can be seen in the composition of DD200 (see Figure\,2),
magnesium is present almost uniformly (in fractional mass) in the outer
layers.  Early in the spectral evolution as the photosphere recedes
through the outer layers, the Mg\,II transition is expected to be
observed with decreasing velocity as measured by the absorption
minimum.  The Doppler shift of Mg\,II will cease changing as
soon as the photosphere has receded below the inner edge in the
distribution (Fig.\,2).  After that phase, the Doppler shift will not
follow the receding photosphere, but remain at an almost constant
velocity.  Thus, early spectra provide a sensitive tool to determine
the boundary between explosive carbon burning and explosive oxygen
burning in velocity space.  In DD200, the velocity of the onset of this
transition is about 15,000--16,000\,km\,s$^{-1}$ and the photosphere
has receded inside this region $\approx $ one week before maximum.
Therefore, the subsequent small change in the line shift in the models
is consistent with the observations (Meikle et al.  1996).  Although
the model agrees reasonably well with the data, the blueshift in the
model is somewhat smaller than in the observations, indicating that the
onset of explosive oxygen burning took place at a slightly smaller
velocity in the model than in the supernova.

In Figure\,7 the LTE model of DD21 is compared with the data at
approximately one week before maximum light. The model produces the
Mg\,II feature although this model exhibits a higher blueshift
than the data.  As can be seen in Figure\,3 the magnesium in DD21 is
truncated at higher velocities than in DD200.

In the DD200 model spectrum a relatively strong line can be seen at
1.15 $\mu$m (Fig.\,6).  This is due to Ca\,II.  In Figure\,2 calcium is
found at the interface layer between the burning to nickel and the
burning to intermediate mass elements.  The absence of the feature in
the data can be attributed to the fact that the model represents a
phase $\sim$ 6 days later than the data when the scattering photosphere has
moved further inwards in the model.  The appearance of this feature is
thus a useful diagnostic of the location of the transition layer
between complete and incomplete burning of silicon.  In SN1994D, a
feature at approximately the right wavelength but without the clear
P-Cygni signature becomes visible about 1--2\,d past maximum (Meikle et
al.  1996).  We note that the optical Ca\,II features
cannot provide this diagnostic. In that wavelength regime, primordial
(not freshly synthesized) calcium contaminates the ejecta and provides
sufficient lines of significant optical depth to confuse the interpretation.
The DD21 LTE model spectrum does not show a strong Ca\,II line
at 1.15 $\mu$m. The difference between the two models can be understood
as being due to the difference in epoch and the difference  in the
structure of the exploding star.

The P-Cygni feature at 1.67 $\mu$m observed in the SN\,1994D data is
reasonably reproduced in both models (Fig.\,7).  In the models, this feature is
due to Si\,II.  Since silicon is produced in all layers above
8,500\,km\,s$^{-1}$, the Doppler shift of Si\,II follows the
photosphere. Clearly the photospheric velocity from the models is not
perfectly matched to the data.  This feature is less likely to be
present in a supernova such as SN\,1991T. In overluminous supernova
such as SN\,1991T less silicon is made and it is hotter and therefore
probably in a higher ionization state.

Both the model and the data are fairly unexciting in the $K$-band
(2.2 $\mu$m) prior to maximum light.  A small feature is present in the
data at 2.05 $\mu$m. A similar, but somewhat weaker, spectral feature is
also present in the LTE model and is due to  Si\,III.
This feature also appears as an opacity source in the non-LTE model, but does
not cause a strong feature because, at the later epoch of the model,
the main ionization stage is Si\,II.

\begin{figure}
\centering
 \plotfiddle{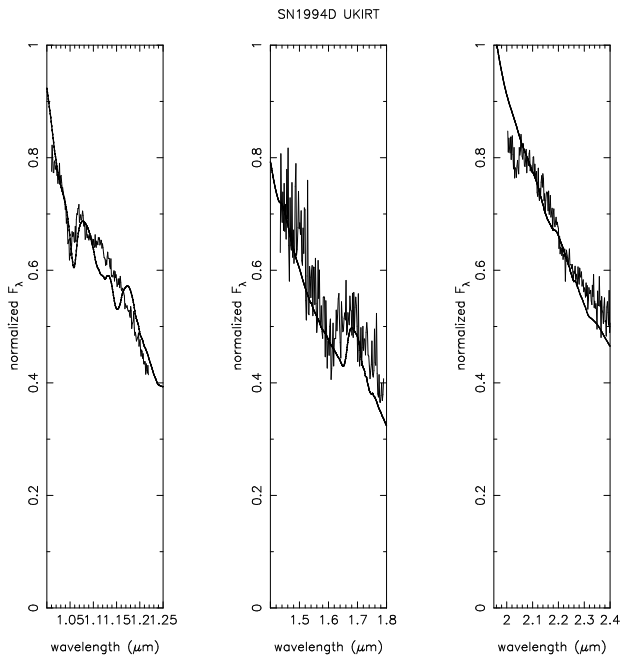}{8cm}{0}{45}{45}{-120}{0}
\caption{Comparison of non-LTE model 2 days before model B
maximum with SN\,1994D observed 8.5 days before B maximum.}
\end{figure}

\begin{figure}
 \centering
  \plotfiddle{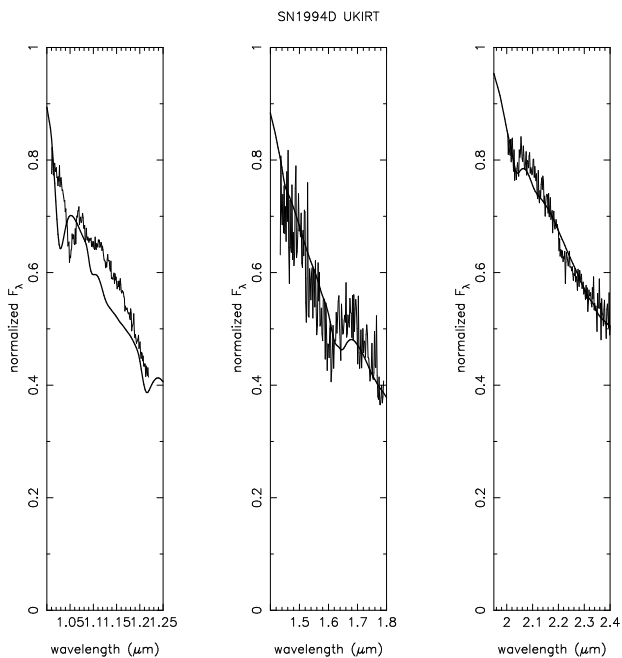}{8cm}{0}{45}{45}{-120}{0}
 \caption{Comparison of LTE model 10 days after the explosion ($\approx$
-6 days before B maximum) with SN\,1994D observed 8.5 days before B maximum.}
\end{figure}

\subsection{Post-maximum Spectra}

\begin{figure}
 \centering
  \plotfiddle{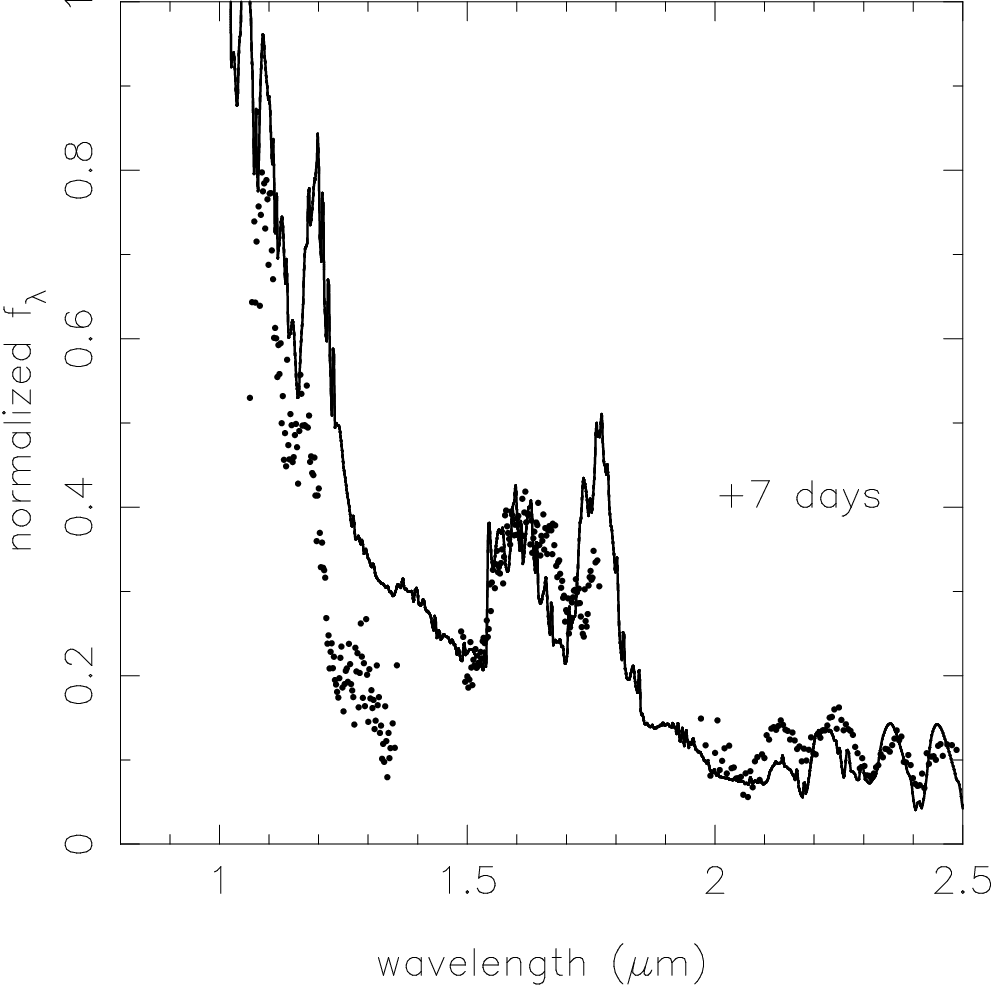}{8cm}{0}{45}{45}{-120}{0}
 \caption{non-LTE model at +7 d and SN\,1986G at +10 d
with respect to model and observed V maximum, respectively.}
\end{figure}

\begin{figure}
 \centering
  \plotfiddle{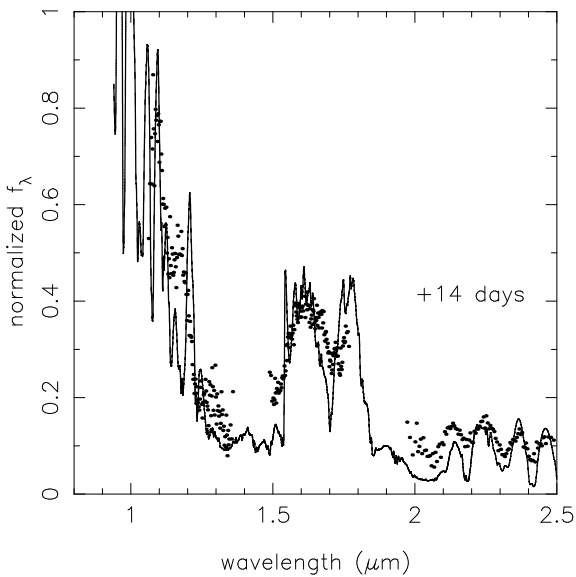}{8cm}{0}{45}{45}{-120}{0}
 \caption{non-LTE model at +14 d and SN\,1986G at +10 d
with respect to model and observed V maximum, respectively.}
\end{figure}

\begin{figure}[t]
 \centering
  \plotfiddle{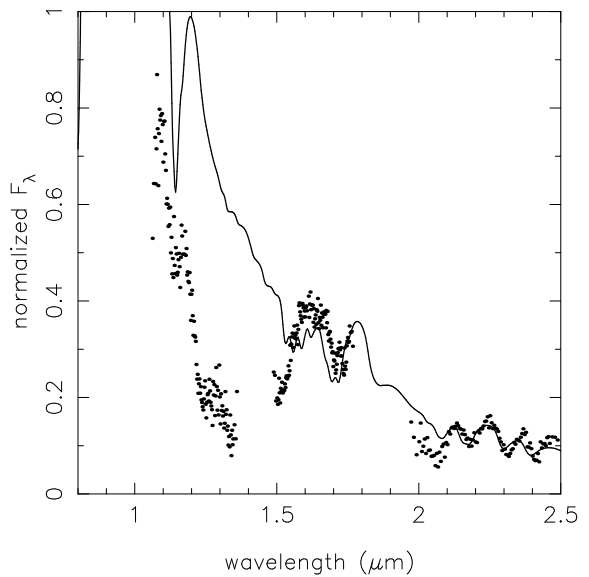}{8cm}{0}{45}{45}{-120}{0}
 \caption{LTE model at 24 days after explosion (
$\approx $ +7\,d after model V maximum) and SN\,1986G at +10d
past V maximum.}
\end{figure}

In Figures\,8---10 we compare the data from SN\,1986G with the models.
Note that SN 1986G is regarded to be a somewhat subluminous, rapidly
declining event compared to ``normal" SN Ia.  We have not
attempted to select models specifically appropriate for these
circumstances, but can address general features and possible
diagnostics of such supernovae.

Overall, the non-LTE models produce
the observed spectral features rather well.  The non-LTE theoretical
spectra in Figures 8 and 9 corresponding to about +7 and 14\,d after
model V maximum light (a day later than model B maximum) bracket the
epoch of observations at 10\,d.  Within the uncertainties, the
differences between the data and the model spectra can be attributed to
the time evolution.  The LTE model 24 days after the explosion corresponds
to about one week after model V maximum light in DD21 (Fig.\,10).
The ratio of the dips in the spectra
at 1.2 and 1.9 $\mu$m and the quasi-continuum in the $H$-band evolve
dramatically in the models.  This can be expected as the continuum
opacity drops.  The iron-group elements are exposed and the
associated strong line blanketing causes the photosphere to form at
large radii (Fig.\,5) while the lack of opacity in the holes leaves the
core exposed.  The LTE model in Figure\,10 does not reproduce the
J-band spectral minimum at $\sim$ 1.2 $\mu$m.  This model gives a
``continuum" at this wavelength and epoch that is caused by line
blanketing due to Si I.
In the non-LTE model,  most of the silicon is in the second ionization stage
and there is very little Si I.
The difference can be attributed to the fact that, in the LTE models, the
matter temperature is given by the energy density and
drops strongly with radius as $T{\propto}R^{-1/2}$.
In the non-LTE models, the matter temperature more closely follows
the radiation temperature (see also Jeffery et al. 1992).
This implies that the  1.2 $\mu$m hole provides a temperature diagnostic
for the silicon layers.

Frogel et al. (1987) suggested that the minimum observed in their IR
spectra of SN\,1986G at 1.7 $\mu$m could be identified as magnesium
with a redshift of --0.0025.  Note that a negative redshift implies
that Frogel et al. used the depressions in the data to identify
the lines.  The narrow dip at 1.7 $\mu$m in both models presented here
is due to a lack of line opacity.  This obvious feature in the data for
SN\,1986G is not visible in the data from SN\,1991T (see Bowers et al.
1997).  As discussed in \S3.2, a plausible explanation for the
difference is that the luminous SN\,1991T has Ni/Co at higher
velocities, thus smearing out the opacity gap at 1.7 $\mu$m.

As mentioned above, the strong quasi-continuum in the $H$-band
(1.6 $\mu$m) is due to a large number of iron group lines.  Although
iron underlies most of the emission, the bluer of the two broad
features in the models is dominated  by cobalt transitions while nickel
contributes strongly  to the redder of the two.  In part, the evolution of
the ratio of the two features can be attributed to the increase of the
cobalt abundance relative to the nickel due to the radioactive decay of
the nickel to cobalt.
The smaller red shift of the flux minimum
at about 1.7 $\mu$m in both models DD200 and DD21 in comparison with
the observations ($\sim$\,1,500\,km\,s$^{-1}$) could be accounted for
by the fact that  SN\,1986G was slightly subluminous (see Phillips et
al. 1987) and had less nickel expanding at a smaller velocity than the
``normal" models compared here.

Frogel et al.  (1987) identified the $K$-band (2 $\mu$m) features as
due to Na\,I with a redshift of --0.00133.  The spectral models
presented here identify these features as being due to transitions of
iron-peak elements and silicon.  The $K$-band features are present in
both the SN\,1986G and SN\,1991T data (Bowers et al.  1997).  They are
well reproduced by both the LTE and non-LTE models.  These features can
be expected to be present in all SN Ia spectra although their
velocities may vary.  In the models, these features are formed in the
transition region between complete and incomplete silicon burning.  The
overlap between these regions is sensitive to the explosion model and
the $K$-band features can be used as a diagnostic of the composition
structure of the two regions that can be clearly seen in Figures\,2 and
3.

Overall, both the LTE and non-LTE models reproduce the strong features
because the basic pattern depends on the atomic physics and the
structure of the explosion model, rather than the level populations.

\section{Discussion and Conclusions}

The near-IR wavelength range has been shown to be a valuable tool to
diagnose explosion models of SN Ia.  The current
generation of atmosphere codes is well suited to allow quantitative
analysis of the observed spectra by means of delayed detonation models.
The delayed detonation scenario works well both for the optical and
IR.  The good reproduction of the IR is an especially satisfactory
result given that the calculations were done without prior knowledge of
the observations and without ``tuning" the dynamic models or the atomic
data, but taking the results as they were.  These models provide a good
basic understanding of the IR spectra.
The IR is thus revelaed as a unique tool providing spectral diagnostics of
basic
quantities such as the boundary between explosive carbon and oxygen and
between complete and incomplete silicon burning by measuring Mg\,
II, Ca\,II and iron group lines.

This paper concentrated on the IR spectral features, but a word
about the IR photometric evolution is also relevant.
Elias et al. presented H-band light curves and the evolution of the J-H
color with time and showed that SN~Ia follow a typical photometric evolution.
J and H peak about 6 days before B maximum (note that all the figures
in Elias et al. are given in terms of time since H maximum, not B
maximum).  The H light curve has a pronounced minimum about 9 days
and J about 14 days after B maximum.  Both J and H have a secondary
flux maximum about 24 days after B maximum.
The J-H color curve reddens dramatically from about 0 near optical maximum
to a maximum of about 1.4 14 days after B maximum.  It then turns blueward
to a value of about 0.3 a little over 30 days past B maximum and
finally moves redward again to a constant value of about 1.7 after 80 days.
Similar non-monotonic light curve features with secondary bumps or inflections
appear in other photometric bands from the V
to the K band with varying intensities (Suntzeff et al. 1998).
Elias et al. argued that the $J$-band light curve minimum resulted
from an absorption rather than a later increase in emission at the time
of the second IR maximum that corresponds approximately to the onset of the
exponential decay in the optical light curve at 30 days past maximum.
Graham (1986) attributed the red J-H color of SN~Ia to the strong ``absorption"
at 1.2 $\mu$m. The current models suggest that Graham was basically correct
about
the overall reddening of J-H. The lack of flux at 1.2 $\mu$m compared to the
increased emission due to the strong quasi-continuum in H certainly
makes the J-H color red.  Graham bolstered his argument, however, by
drawing a parallel between the behavior of the J-H color curve and
the strength of absorption of the Na D line.  The latter may not be
reliable because in the relevant time frame, the [Co III] line is
suspected to begin to yield net emission at just the wavelength of
the Na line.  In any case lacking a model, Graham did not actually attempt
to account for the non-monotonic behavior of the J and H band
light curves with their distinct secondary maxima.

H\"oflich, Khokhlov, \& Wheeler (1995) have presented an explanation
for the broad band non-monotonic behavior of the IR light curves in which
time dependent opacity can lead to an effective emitting radius and hence area
that peaks at a delayed time in the IR compared to the optical.  Thus their
explanation is just the opposite of that of Elias et al. (1985).
The non-monotonic IR light curves are due to a later increase in flux,
not an intermediate ``absorption.  In
these models, the 1.2 $\mu$m spectral ``hole" does not
play a direct role in the light curve evolution.  The  1.2 $\mu$m spectral
feature is undoubtedly the basic source of the very red colors of SN~Ia after
maximum, but its presence does not yield a direct explanation for the
non-monotonic behavior of either the J band light curve nor the
evolution of the J-H color.  This issue is surely not yet closed.  We note
that despite their success, the models of H\"oflich et al. predict the
first peak in J and H to occur quite close to B maximum, rather than
6 days earlier as the data seem to suggest.

Finally, we  want to stress that this paper should not be regarded as a
final answer, but a first step toward the analysis of the IR spectra of Type Ia
supernovae. To utilize the full potential of the IR, detailed analysis
of IR spectra together with optical spectra and light curves are
essential to answer questions concerning the structure of Type
Ia supernova.

\subsection*{ACKNOWLEDGMENTS}

James Graham, Peter Meikle and Emma Bowers are thanked for providing
data in digital form and Eddie Baron, David Branch, Peter Meikle,
Ken Nomoto, and Douglas Swartz for comments on the manuscript.
JS gratefully acknowledges the hospitality of
the University of Texas at Austin, where this work was started.
JCW, PAH and JS would like to thank the Institute of Theoretical Physics
at the  University of California at Santa Barbara, where this work was
finished.
The ITP is supported by the NSF under Grant No. PHY94-07194.
This research was supported in part by NSF Grant AST 9528110,
NASA Grant NAG 5-2888, and a grant from the Texas Advanced Research Program,

\end{document}